# The Compute ICE-AGE

*OPAL ONE: continuity that persists while reasoning is idle*

**R. Jay Martin**
Lake Arrowhead, CA | ray@opalone.ai

*Abstract*

We built OPAL to give applications memory that lasts between reasoning sessions. The engine runs on the CPU, keeps records and their relationships, applies updates, and recovers saved work after a restart. KINDRED/MCP and PYAMV put that engine to work through application and developer interfaces. We tested selected operation latency across retained-collection sizes and separately tested defined reliability contracts. BLGC supplies the mathematical foundation. This paper reports the working engine, its tested interfaces and measured results.

We measured updates to existing records and events that carry an identity so the engine can recognize a retry. In a same-session sequence at 1M, 5M and 1M retained nodes, the current engine recorded resident-update P50 of 0.083 microseconds at all three sizes/order positions. Every process passed its correctness checks. The fixture used 1,024 prepared targets and the same frozen binary. Earlier five-process 5M measurements recorded subsequent receipt P50 of 4.708-5.166 microseconds on the preceding engine version; those results retain their original version and sampling scope [8, 13].

We tested reliability alongside speed. Separate tests cover restart and continuation, event identities, rejected requests, simultaneous calls and memory-allocation failures. Comparisons across engine versions produced exactly matching checkpoint files and matching state after recovery and further work. Storage-growth tests show the occasional costs of expanding containers. Appendix B keeps the historical comparison and repair details. Ordinary local operations remain inexpensive through five million nodes in these tests.

We built a PYAMV installation package, or wheel, and installed it in a clean Python environment. It passed 17 SDK checks and all 20 workload checks across five repetitions of four public-SDK workloads. KINDRED connects models to project memory through MCP; PYAMV lets developers call the engine from Python and repeat the benchmark workloads. KINDRED also passed 12 representative MCP checks and ten additional authorization and transport checks. We plan to release both under free licenses for local experimentation by engineers and home builders.

The Compute ICE-AGE names the separation at the heart of this system: retain continuity in memory, update it through local operations, and invoke reasoning when the task requires it. The original operating record reported a cool-running CPU near 17.2% utilization; the new studies measure latency and correctness rather than temperature or power. The next step puts these capabilities in more hands. Engineers can build local tools; EMERGENCE takes the work toward personal AI that belongs to its user.

# 1. Continuity While Reasoning Is Idle

**OPAL gives an application memory that lasts while reasoning is idle.** It keeps records, their identities and their relationships. It applies updates under defined rules, saves the result, and recovers it when a new process starts. The model interprets, plans and writes. OPAL keeps the record between those activities. Throughout this paper, “state” means the information the engine currently holds; “accepted” means an operation passed the engine's rules.

For someone returning to a project, continuity means more than seeing an old conversation. It means keeping an accepted decision distinct from a suggestion, retaining a correction alongside the information it supersedes, and carrying unfinished work forward. KINDRED connects that memory to project information, corrections, handoffs and whiteboard tasks. PYAMV exposes native operations to Python programs. These interfaces give the engine a practical role in work that continues across interactions.

We wrote OPAL in C++17 to run on the CPU. The engine can maintain memory without asking a model to generate an answer. Applications keep their records while reasoning pauses and use those records when work resumes. OPAL ONE builds on that separation. EMERGENCE takes it toward personal AI under the user's control.

Applications often gather history, documents and earlier answers every time they call a model. As history grows, that preparation can move more data and demand more computation. Retrieval-augmented generation, or RAG, selects material for the prompt. The application still needs to keep its decisions, identities and progress between calls. OPAL gives that record a lasting home.

A probabilistic model generates a response $s_t$ using its parameters θ and earlier context $C_{(<t)}$:

$$s_t \sim P(s \mid C_{(<t)}, \theta).$$

That expression describes how a model generates a response. It does not specify where an application keeps its record between responses. OPAL stores that record under stable identities, so the application can find and update it later. The model reasons; the engine remembers.

We built two wrappers to use and test the engine. KINDRED provides a Model Context Protocol (MCP) server and an application that routes model requests and tools. PYAMV provides a Python software development kit (SDK). KINDRED connects memory to project work; PYAMV lets developers write their own programs and experiments. The native tests measure engine operations directly. The installed-SDK tests measure calls through Python. These are working interfaces, with distinct jobs and distinct evidence.

The measurements establish the behavior of the working engine. The current same-session 1M, 5M, 1M sequence records resident-update medians of 0.083 microseconds throughout. The earlier receipt series records medians of 4.708-5.166 microseconds at 5M on its recorded engine version. Separate preservation and recovery checks establish reliability for their tested rules. These results demonstrate the engine's local update performance, state preservation and recovery.

**The Compute ICE-AGE is a shift in how a system spends its computation.** It retains accepted continuity, performs local updates where their contracts allow, and prepares relevant state for reasoning. The reported cool-running CPU motivates the name. The following sections explain what the engine does, present the evidence from native operations and its Python interface, and connect those results to personal and hosted applications.

# 2. Memory That Persists

OPAL provides the memory foundation for OPAL ONE. An application hosts the C++17 library and calls its operations. A model can reason over information from that memory while OPAL keeps the records and their relationships between calls.

We represent the stored information as a graph, $G = (V, E)$. Nodes hold records under stable identities; edges connect related records. "Semantic continuity" here means keeping those identities, connections and accepted values. We test whether the engine preserves them separately from whether a model interprets them well.

The application decides what information to submit. OPAL checks the request and applies the relevant update rules. The application can then find, update, save and recover that record.

The rules distinguish reading from changing memory. A passive read leaves state alone. An explicitly reinforcing access can strengthen it. Each query specifies which behavior it uses. Stable identities let applications find records across calls, and saved checkpoints let a new process recover the record and continue.

An index lookup and a graph traversal do different work. An index helps an update find a particular record. A traversal follows relationships. The local update need not search the whole graph or rank similar vectors. The resident operations measured here run on ordinary CPU hardware without a GPU model call.

OPAL saves graph state through CBOR, a binary data format. Earlier save-load tests produced matching graph hashes even with noisy input and disrupted relationships. The newer comparison goes further: the original and repaired versions produced byte-identical checkpoint files, including after fresh processes recovered them and continued the work (Section 7). A checkpoint is a saved state from which the engine can restart; canonical files use the defined encoding that allows exact comparison.

This separation gives applications room to choose. A personal assistant and a server-hosted agent can use the same memory foundation while choosing different models and tools. Each application controls how it selects information and accepts updates. Section 10 explains those uses. Section 3 explains what keeps a particular update local.

# 3. Bounded Local Evolution

Bounded Local Generator Classes for Deterministic State Evolution supplies the mathematics behind the design [1]. BLGC specifies conditions under which an update needs a bounded amount of work as the total number of records grows: bounded local state, controlled dependencies, suitable update rules, and bounded work to obtain and evaluate the inputs. The revised manuscript uses $\ell^\infty$, the space of uniformly bounded configurations [2]. That companion paper contains the proofs.

Time affects the update. Elapsed intervals control eligible changes; chronology rules decide whether the engine accepts an event and continues from it. A revision counter does not measure elapsed time. The schedule and checked context determine which operation applies.

An ordered operation changes the graph from one state to the next:

$$G_{t+1} = g_t(G_t).$$

This notation includes the checked context and the commit that accepts the change. It does not equate every stored field with a coordinate in the mathematical model. Evaluating a fixed-size numerical update and completing the whole transaction impose separate costs.

With fixed numerical dimension, a bounded number of inputs, and bounded work to obtain and evaluate them, BLGC gives:

$$W_{local} = O(1) \text{ with respect to } M = |V|.$$

Limiting traversal distance alone does not limit the number of edges an operation examines. The operation must also control how many connections it considers, how it selects them and how it accesses them. The resident update traced here uses one target with two numerical coordinates and fixed context. Those coordinates accumulate accepted input; BLGC's projected unit-ball coordinates use a different domain. The companion study explains the input bounds over a finite horizon and the distinction between the production and formal state domains [2]. We kept the engine's update law unchanged.

The aim is straightforward: keep more records without requiring each local update to examine more inputs. We test that relationship by timing the actual operation.

The public-call measurements include finding inputs, evaluating the update and committing it. Appendix B keeps the comparison and audit that explain this work.

## From Formal Conditions to Executable Tests

BLGC proves properties of a specified generator class under explicit hypotheses. This study tests how selected production operations realize the local-work design and preserve accepted state. The bridge is concrete: identify the operation and its inputs, hold the workload shape fixed, increase retained state, and check both work and resulting state.

| Formal or systems obligation | Implementation and experimental check |
|---|---|
| Fixed local numerical work | Trace the two-coordinate resident transition and fixed context; hold dimension and degree fixed across 10k-5M. |
| Bounded acquisition for the tested path | Measure the complete indexed public call, including validation, locking, lookup and commit; account separately for initialization and storage growth. |
| Defined state evolution | Compare the production transition with reference-state calculations; check chronology, generation and replay behavior. |
| Durable accepted state | Compare canonical checkpoints and independent-process recovery/continuation; exercise failures and concurrent identities. |
| Usable developer interface | Build and install PYAMV, check its public contracts, and run validated lookup, ingestion, traversal and recall workloads. |

BLGC supplies the conditional proofs. This paper supplies production measurements and correctness checks for the named operations, with the formal and production state domains distinguished above.

# 4. The Calculus Engine

Memory must handle change. Noisy input, uncertain model answers and changing relationships can leave a graph with isolated records, weak connections and unstable paths. Keeping the data alone does not keep the structure useful.

The calculus engine regulates change through reinforcement, decay, alignment and compression. It works on the graph itself. The implementation includes stabilizeEntropy(), evolveRelevance(), relevanceCurve(), propagation decay functions and centroid alignment scoring. These mechanisms come from the original design. The resident-event measurements below do not assign the same work bound to every calculus operation or to global Flow maintenance.

## How the Engine Regulates Memory

The engine adjusts relevance using traversal frequency, reinforcement from relationships, agreement among related records, propagation stability and the strength of local connections. Repeated use can reinforce a well-connected group. Weak or inconsistent connections can lose influence and leave a record isolated.

Input can remain uncertain while the storage rules remain deterministic. OCR errors, model output, noisy numerical representations and broken connections can all enter the system. The calculus layer applies defined rules to that changing graph. Deterministic storage preserves what the engine accepts; it does not make every input correct.

## How the Operations Work

Entropy stabilization limits the spread of influence across weakly related regions. When noisy input or heavy traversal disrupts connections, local routines constrain propagation and isolate unstable neighborhoods. stabilizeEntropy() and the propagation decay functions perform this work within bounded traversal neighborhoods.

Relevance evolution changes a record's influence over time. Traversal, reinforcement, agreement among related records, local connections and propagation stability affect the result. evolveRelevance() and relevanceCurve() implement those changes. Frequently used, coherent groups can strengthen; weakly connected groups can lose influence.

Centroid alignment scoring measures how records align around the center of a local group. Traversal and updates strengthen coherent regions while unstable paths weaken. The original design uses this local alignment to limit fragmentation and keep traversal stable as the graph changes.

Propagation decay reduces influence as traversal proceeds. Depth, coherence and structural reinforcement affect the reduction. This limits how far an update spreads through unrelated regions.

Compression reduces structure through local changes, reinforcement and relevance rules. The design combines storage reduction with relationship preservation and exact save-load recovery. Its costs depend on the records and connections each operation examines.

Global snapshots, geometry calculations and policy-history processing perform additional work. We keep their costs separate from the fixed-context events measured in this paper.

## What the Stress Tests Showed

We deliberately challenged OPAL with corrupted text, uncertain model output, malformed graphs, fragmented connections, random propagation and missing relationships. These workloads tested state preservation with noisy input and damaged graph structure.

The historical tests recorded fragmentation and orphan records while replay checks continued to pass. Save-and-recovery comparisons matched the stored graph fields that those checks covered, including the damaged structure. The newer tests separately verified accepted updates, replay handling, rejected conflicts and continuation after restart.

These results make reliability concrete: OPAL preserved the tested record even when the application supplied imperfect information. Preserving that record and improving its relationships are distinct outcomes.

# 5. Experimental Environment

We ran the main experiments on an Apple Silicon M2 Pro with 16 GB unified memory. OPAL uses C++17 and saves CBOR data to local storage. The original paper reports sustained workloads and an iPhone simulator integration. The newer tests call the production engine APIs directly.

## Workload and Data Structures

The earlier operating record reports 1M-25M-node workloads with orderly and deliberately damaging synthetic input: OCR-derived corruption, malformed graph structure, noisy embeddings, fragmented connections, random propagation and ANN/LSH disruption. This paper retains those observations as historical evidence. They are distinct from the new fixed-parameter series and from the separately preserved six-run 1M/5M workload [3, 4].

The controlled family uses a directed ring with two successor links per data node, fixed record shape, 13-byte IDs, eight-component embeddings, a two-coordinate numerical state and 1,024 evenly spread targets in a fixed permutation. Policy, stored context and event schedule remain fixed. We hold the local connection pattern and degree fixed as total node count changes.

## Measurement Methodology

The earlier production series uses the engine version documented in Appendix B, in native arm64 Release with Apple clang 21.0.0, C++17 and -O3 on macOS 26.6.2. The 22 October series supplies three fresh engine processes at each of 10,000, 100,000 and 1,000,000 nodes. The 23 October extension supplies five at 5,000,000 nodes using the same frozen executable. One caller performs the operations; OPAL_THREAD_COUNT is one. No profiler was active. Appendix B records the accompanying baseline runs and matched execution order [5, 8].

We record setup, initial index construction and warmup separately from the primary timings. Each process measures 20,480 passive snapshots and 20,480 resident updates. The update finds the target, validates the request, evaluates the time-dependent change, commits it, and updates the generation counter and cache. Explicit one-second intervals and alternating small signed contributions exercise real transitions. Outside the timed calls, the test checks every target against an independent reference. It also checks chronology, content, relationships and an unvisited node.

Each process then measures 16 receipt events with stable identities. The first begins with a prepared index; the remaining 15 form the subsequent-receipt sample. Count increases from M to M+16. Steady-clock per-call samples use nearest-rank quantiles; whole-loop timing is separate. The records retain empty timer-pair diagnostics. The nine engine processes at 10k-1M contribute 184,320 calls per primary operation and 144 receipt events [5].

A separate nine-process fixture forces vector and hash growth, three runs per scale, and checks the resulting state and index. Separate processes test recovery. The author-held evidence packages retain frozen sources, binary identities, execution order, raw samples, canonical comparison files and validation logs [4, 5].

The five 5M engine processes add 102,400 passive snapshots, 102,400 resident updates and 80 receipt events. They preserve the fixture, executable identity and correctness checks. The host reported 3,311.75 MiB of swap in use before execution. This is a shared-machine follow-up, not a new counterbalanced rerun of all scales [8].

The new series did not collect CPU utilization, temperature, fan speed or power. Section 8 addresses the original sustained-operation observations. We do not use latency to measure temperature.

# 6. Core Measurements

The engine performs ordinary local operations at low cost within a five-million-node collection. Here, P50 means the median; P95 and P99 mark the values at or below which 95% and 99% of samples fall. A fresh process starts a new program instance for the run. The following results report the production path, its correctness checks and its timed steps. The earlier scales provide context; Appendix B retains the implementation comparison.

## Production Operations at Five Million Nodes

**The current engine records resident-update P50 of 0.083 microseconds throughout a same-session 1M, 5M, 1M sequence.** All three processes pass their correctness checks [13]. Each process uses the same frozen binary, 1,024 prepared targets, warmup and timed public-call boundary. The timer-pair P95 is 42 ns; equal medians establish stability at this method's resolution.

A generation counter identifies state changes. Chronology checks enforce event order. Replay recognizes a matching event identity; a conflict means the same identity arrives with different content.

A resident update changes the numerical state of an existing record in memory. A receipt event also records the identity of the accepted event, allowing the engine to recognize a matching retry and avoid applying it twice. Both timings cover the complete public call in memory: checking the request, acquiring the lock and making the accepted change visible. Receipt timing also includes constructing and appending the event payload. Neither measures disk checkpoint publication or traversal of the entire graph.

**Table 1.** Current-engine resident updates [13]. One sequential process at each position in the same-session sequence. P50 uses 20,480 calls per process. Mean and whole-loop time retain additional information beyond the quantized median. All values are microseconds.

| Collection / sequence position | Resident P50 | Mean | Whole-loop per call |
|---|---|---|---|
| 1M / first | 0.083 | 0.073349 | 0.092159 |
| 5M | 0.083 | 0.077172 | 0.099143 |
| 1M / return | 0.083 | 0.073988 | 0.093062 |

The current-engine validation also exercised 8,192 IDs arranged as 4,096 deliberate collision pairs. It passed 40,960 validated updates, with P50 0.083, mean 0.084554, P95 0.125 and P99 0.167 microseconds. Compared with the reference binary, the deliberate-miss workload added 8.828 ns to the mean and one 42 ns timer quantum to P99; P50 and P95 matched. Independent recovery and continuation checks passed, and canonical checkpoint and continuation files matched byte for byte [13].

The following 10k-5M methodology, receipt measurements and memory figures describe the earlier series [5, 8]. Appendix B retains the receipt comparison and timing audit. These records and the current-engine sequence are separate experiments; the new resident result does not replace the earlier receipt measurements.

The workload holds degree, record shape, numerical dimension, policy and target schedule fixed. It exercises 1,024 prepared targets within the retained collection. Setup, initial index construction, warmup and final verification remain outside the individual-call timers. The 5M series uses the same frozen executable as the smaller-scale series, but runs in a later shared-machine session with preexisting swap use.

We separate repeated-call statistics from single first-call observations according to their different sampling methods. The earlier series reports resident-update P50 over 20,480 calls and receipt P50 over calls 2-16; Appendix B retains both summaries. Appendix B reports the single first receipt observation from each run because one observation per process cannot define a latency distribution. This presentation retains every measured result. With only 15 subsequent receipt samples per run, the receipt summaries characterize the exercised calls but do not estimate production tails.

At 5M, peak resident set size (RSS), the operating system's resident-memory measure, for the tested process ranges from 4.377 to 4.638 GB (decimal GB, macOS high-water accounting). Existing swap use does not establish per-call paging. Empty timer-pair diagnostics in the initial study recorded P50 of zero and P95 of 42 ns, limiting interpretation of small differences in resident timing [4, 8].

Separate validation fixtures test checkpoint recovery, concurrent identities and allocation failures; the 5M runs retain their scale-correctness scope. Appendix B preserves the pre-repair comparison, source-level explanation and timing audit. Those records document where the evidence came from without changing the operation definitions or results.

## Python SDK: Using the Engine Through an Installed Package

PYAMV exposes native engine operations to Python. On 26 October, we built a Release wheel from the current source, installed it into a clean Python 3.14.6 environment on macOS arm64, and verified that the workload loaded the installed package. All 17 SDK contract checks passed. Five sequential repetitions of the existing local DCPerf-oriented, Benchpress-compatible workload each passed all four operation-specific checks: 20 successful workload processes in total [10].

**Table 2.** Installed-wheel SDK throughput. Median and observed range across five run-level rates. Each workload uses 100 warmups and 1,000 measured calls in a fresh child process; rates include Python, harness and validation overhead.

| SDK operation | Median rate | Observed range |
|---|---|---|
| Indexed lookup, calls/s | 365,820 | 355,909-370,222 |
| Text ingestion, calls/s | 27,245 | 26,518-27,897 |
| Text ingestion, states/s | 217,956 | 212,147-223,174 |
| Connected traversal, calls/s | 65,394 | 61,855-67,888 |
| Direct node-ID recall, queries/s | 59,681 | 59,443-60,318 |

Lookup uses 128 existing states; direct node-ID recall uses 96. Connected traversal uses a 1,200-state fixture and a two-hop request. The workload checks reachable starting points, application-cache misses and valid paths. Each measured text-ingestion call inserts eight states and checks the expected state and relationship additions. Recall checks the requested identity, not semantic answer quality. A whole-loop timer supplies each throughput rate; per-call latency samples remain diagnostic.

The installed SDK retrieves records, adds states and relationships, follows connected paths and recalls identities through its public API. We measured those operations through the developer interface. These are smaller operation-specific fixtures, separate from the native 5M scaling study. We ran the Benchpress-compatible SDK series ourselves. We did not run the upstream Meta DCPerf suite, and these results do not claim independent third-party testing. Source hashes, wheel identity, installation logs, checks and raw reports accompany the evidence [10]. The engine and workload source remained unchanged during the experiment.

## Traversal Stability and Earlier Scale Evidence

The earlier operating record reports local traversal through 1M-25M-node workloads, with greater timing variation under paging and damaged graph structure. The preserved historical figures remain in Appendix A. We keep those observations separate from the new resident-event timings because they measure different operations.

The separately documented six-run series completed three fresh workloads at one million nodes and three at five million. Indexed reinforcing access used one warmup and 100 timed calls for each of 100 deterministic IDs, with percentiles across per-ID means. P50 ranged from 0.12541 to 0.12874 microseconds. These are indexed-access statistics; they are not individual-call traversal quantiles [3].

That series also measured persistence, Flow refresh, integrity and traversal checks. Maximum sampled whole-workload RSS ranged from 3.025043 to 4.065346 GB at 1M and 6.215483 to 8.420590 GB at 5M. These are discrete whole-process observations under larger-record workloads, not the footprint of the new small-record fixture.

## Memory Pressure and Paging

The historical experiments deliberately grew the graph beyond physical RAM during sustained traversal and updates. The operating system then paged data between RAM and storage. Those tests exposed how memory pressure affected the engine.

Paging moved portions of the graph into SSD-backed virtual memory. Traversal took longer and varied more because the system had to retrieve pages from storage. Sustained updates also had less predictable access to nearby records.

The historical record reports gradual degradation, local fragmentation and passing replay checks while the graph exceeded RAM. The new controlled series does not independently reproduce the 25M or sustained-paging regimes. It exposes a different, directly attributable source of scale-dependent latency: capacity growth.

**Table 3.** Forced storage-growth event durations, in milliseconds, across three fresh processes per scale. One event of each type per process; untimed preparation places the container at its growth threshold.

| Data nodes | Vector-growth event | Hash-growth event |
|---|---|---|
| 10,000 | 0.760-1.136 | 0.140-0.155 |
| 100,000 | 7.439-9.133 | 1.167-1.511 |
| 1,000,000 | 189.330-251.888 | 70.486-79.657 |

All nine forced-growth processes passed state, index and replay checks. Incremental indexing removes repeated reconstruction from ordinary receipts; occasional allocation and container growth retain their own costs. Other receipt-producing operations, policy-history scans and cache maintenance remain outside this tested path.

## Replay Integrity and Persistence Stability

The replay tests saved the graph in CBOR and reconstructed it. Each check calculated a graph hash before saving and compared it with the hash after recovery. A hash provides a fingerprint of the fields that the check covers.

The earlier graph-hash checks address the fields included in their fingerprint. The new comparison provides stronger evidence for the tested repair: baseline and repaired canonical checkpoint files are byte-identical, including the saved state after fresh-process recovery and continuation. These checks directly test preservation across the implementation change.

### Tests with Approximate Similarity Links

The historical tests disrupted approximate nearest-neighbor (ANN) and locality-sensitive hashing (LSH) links. These methods use numerical similarity to select related records. The tests introduced random neighborhood changes, fragmented connections and noisy embedding relationships during sustained operation. An embedding represents information as a numerical vector.

As disruption increased, traversal became less local and paths became less stable. Groups fragmented, orphan records appeared more often, and connections weakened. Replay hashes still matched in the reported tests.

That distinction matters: poor relationships can reduce retrieval quality even when storage preserves the graph correctly. The new event comparison does not repeat these experiments or measure retrieval quality. We retain the earlier observations as part of the operating record.

## 7. Reliability and Failure Experiments

Reliability makes continuity usable. A system must recover accepted state, recognize a repeated event and reject conflicting or invalid requests under its defined rules. The following checks test those behaviors directly, then examine the historical damaging-input findings.

### State, Replay and Recovery

**The engine passes explicit state, event and recovery checks alongside its performance measurements.** All five 5M engine processes pass their declared scale-correctness checks; the accompanying five baseline processes also pass [8, Appendix B]. The study checked reliability separately from timing, using behavior through the public API and canonical state comparisons.

**Table 4.** Reliability rules and observed outcomes [5]. Appendix B records the version comparison and implementation details.

| Rule | Observed outcome |
|---|---|
| Canonical state preservation | Baseline and repaired checkpoint bytes matched exactly; the differential operation transcript also matched. |
| Restart and continuation | Independent processes recovered the checkpoint, replayed the saved identity and continued; resulting saved state matched byte for byte. |
| Event identity and usage | Commit, replay and conflict behavior passed; legacy v2/v3 receipt recognition and usage-once behavior passed. |
| Concurrent same-identity calls | Eight callers produced one commit and seven replays, with one generation increment. |
| Invalid input | The engine rejected stale chronology, nonfinite input and policy mismatch; the covered state and generation checks passed. |
| Lookup and target semantics | Short-term targets, first-duplicate precedence and long-term-before-short-term resolution passed. |
| Index allocation failure | Injected entry/bucket allocation failures left a dirty, rebuildable cache; normal lookup restored it. |
| Append and capacity boundaries | Vector allocation-failure and growth checks passed; 200 receipt appends retained a complete clean index. |
| Checkpoint failure and retry | Failed publication followed by retry retained identical canonical bytes. |
| Scale correctness | All 18 matched and nine forced-growth processes passed their declared state, index and replay checks. |

Appendix B shows how the two versions preserve behavior and how the index recovers after allocation failure.

The broader evidence includes four fresh 513-node runs with concurrent queries: 32,000 successful complete queries while the engine was idle or performed temporal events, Flow refreshes and archive restoration. Relationship-membership and eligibility checks passed. The reported overlap is with each complete maintenance operation, not proof of zero lock contention [6].

These results establish reliable behavior for the exercised operations and failures. They do not derive reliability from a low latency number: state preservation, correct rejection, exactly-once behavior for the tested path that tracks event identities, and restart continuation have their own evidence.

## KINDRED: Project Continuity and Access Controls

KINDRED puts OPAL to work in an ongoing project. Its six public MCP tools - context, recall, remember, correct, handoff and whiteboard - let an application carry accepted facts, corrections, unfinished work and tasks between interactions. All 22 unique focused MCP checks passed [11]. They exercised durable writes that recognize retries, correction history, handoffs, whiteboard isolation and recovery, and the controls that keep each request within its authorized project.

Those protections matter when work continues. A repeated request must not apply the same change twice. A correction must retain its history. A request without valid project access must stop before it reaches the backend. The focused checks verified these behaviors, along with transport limits, request cleanup and diagnostic redaction. They give the continuity interface evidence beyond engine timing alone.

Finder extends that work to user-approved files. KINDRED catalogs linked project roots, tracks stable file and version identities, detects changes, and imports supported text into the owning project's OPAL runtime. Durable import receipts support retries and conflict detection. Bounded snippets carry source identity and distinguish current files from historical versions. The model cannot use an arbitrary path to bypass the approved roots. Finder supplies this evidence behind the six-tool interface.

The current MCP checks and the historical Finder checks have separate evidence records. Appendix C preserves their counts, scope, repair details and remaining limits. Together, the implemented interfaces connect persistent memory to the project work people return to.

## State Preservation Under Damaging Input

The historical stress workloads combined malformed graphs, random propagation, broken connections, ANN/LSH disruption, noisy numerical representations, OCR errors and memory pressure. They recorded isolated nodes, weaker connections and less stable traversal paths as disruption increased.

The 8,004-node replay tests give a concrete result. Deterministic input produced zero orphan records and matching graph hashes. Equivalent workloads with random or noisy input produced 39 orphan records, and the replay checks also returned PASS. The engine preserved the tested stored structure through save and recovery.

The orphan counts describe graph structure; the matching hashes establish preservation of the fields that the checks covered. These observations document behavior under deliberately damaging workloads. The newer reliability tests add direct evidence for update, retry, conflict and restart behavior.

# 8. What the Compute ICE-AGE Means

The Compute ICE-AGE describes where the system spends its work. OPAL keeps the record between reasoning sessions. Local operations update it under defined rules. A model reasons over the information the application supplies. The record can grow without forcing each ordinary update to rebuild the whole collection.

## CPU Behavior During Sustained Work

The original operating record reported a cool-running CPU during sustained multi-hour traversal and mutation, with approximately 17.2% utilization under normal desktop conditions and no observed steady rise as the number of graph nodes increased. That observation motivates the ICE-AGE name: keeping more records did not appear to make the memory engine run continuously hotter.

We retain this as an author-reported operating observation. CPU utilization depends on the monitor's accounting convention and does not measure temperature or power. The available evidence contains no synchronized temperature, package-power, fan-speed or ambient-condition series for those historical runs. The new latency experiments also omitted thermal and power instrumentation. We report the cool-running observation separately from the measured latency and correctness results.

## Where the Work Happens

The production results give the premise a measured foundation. A resident update consumes one target and fixed context. Current resident-update P50 is 0.083 microseconds at 100,000, 1,000,000 and 5,000,000 nodes. The same-session 1M, 5M, 1M sequence is reported in Table 1 [13]. The earlier five-process 5M series recorded subsequent receipt medians of 4.708-5.166 microseconds [8]. A receipt with a stable identity adds the work of committing the event and recognizing retries. The engine performs these operations within the retained collection without calling a reasoning model.

Those measurements establish ordinary execution cost. The separate reliability fixtures test preservation and recovery, while the forced-growth series exposes larger bursts from allocation and container growth. Together they document execution cost, state preservation, recovery and storage-growth behavior.

## Accounting for the Whole Operation

We account for the work in three parts:

$$W_{event} = W_{acquire} + W_{local} + W_{commit}.$$

The event finds its inputs, calculates the local update and commits the result. Even unchanged numerical state can require validation, lookup and publication work. The original diagrams connect model size, the amount of context, neighborhood size and the size of a change. They explain the design; they do not assign a universal execution cost to those relationships.

Total system work also includes ingestion, storage growth, policy/history processing, global geometry, persistence, queries and reasoning. BLGC constrains a specified transition when its acquisition and evaluation assumptions hold. The production measurements respect those boundaries; Appendix B records the implementation audit.

**ICE-AGE means retaining continuity with low recurring work where the operation permits it.** Reasoning can remain idle between calls while accepted state endures; the host still incurs storage and maintenance costs. As retained capacity grows, memory density and locality become important constraints. The next section examines that storage side of the architecture.

# 9. Compression Economics

Once an application retains continuity, storage density and locality become practical design constraints. The engine must hold identities, relationships and state within a usable memory budget. The following historical density observations explain that side of the ICE-AGE proposition and distinguish serialized capacity from runtime memory.

## Per-Node Density Metrics

The original paper reported approximately 1.3 KB per node in the saved Float64 format, with a 128-component embedding. The embedding accounts for 1,024 bytes, approximately 79% of that serialized footprint. The original paper reported approximately 687 bytes per node for a compressed Float32 configuration. Float64 and Float32 store floating-point numbers at different precisions. These remain historical storage observations; they are not the eight-component fixture used in the new timing study.

Embedding precision can dominate serialized size for that record shape. Runtime memory also includes indexes, allocator overhead, relationships, receipts, policy history, caches and temporary maintenance state. Variable degree and content can change those costs. A file-size measurement therefore does not tell us how much RAM each node needs.

## Scaling to Billion-Node Regimes

At 687 bytes per node, dividing 1 TiB by the serialized footprint gives approximately 1.6 billion node records. This is a storage-capacity calculation, not a demonstrated in-memory deployment. The original Int8 projection of more than 2.5 billion records per TiB likewise remains a density scenario rather than a measured runtime result.

These density calculations concern saved records. Numerical precision and distributed access, ordering and recovery determine the behavior of a particular deployment; the capacity arithmetic does not measure those behaviors.

## What This Means for Local Devices

Rebuilding context can consume bandwidth, working memory and model computation even when little has changed. That recurring work matters on a device with limited power or connectivity.

OPAL lets an application keep its record locally on ordinary CPU hardware. The application can maintain that record between reasoning sessions and choose when to call a model. This moves the continuity problem into memory management.

The financial and energy benefits depend on the workload, memory use and frequency of model calls. This study supplies no energy-price or deployment-cost comparison. It measures engine operations and state preservation; energy prices and deployment costs sit outside those measurements.

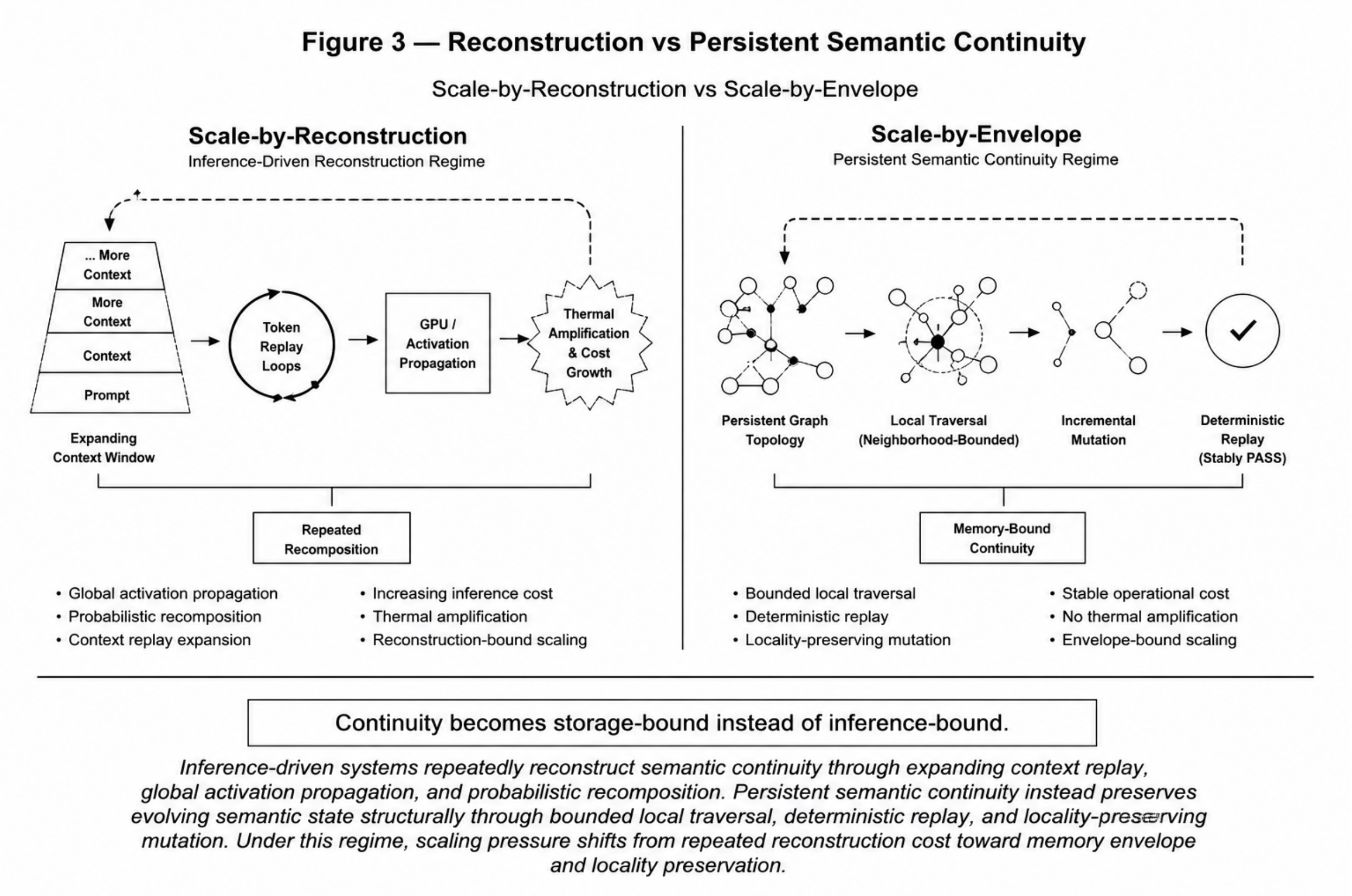

**Figure 3. Original architectural illustration: repeated reconstruction versus growth within a memory budget.** We retain this figure from the original manuscript. The labels express the design thesis; "no thermal amplification" and "stable operational cost" are not universal measured guarantees. Sections 6 and 8 state the measured operation boundaries and thermal evidence.

# 10. What People Can Build

The results connect three practical requirements: keep the record, update it efficiently and fit it in memory. OPAL ONE puts those requirements beneath the reasoning model. The model interprets, plans and writes; the engine keeps the record.

## Three Products, One Foundation

OPAL provides the shared memory foundation. Three distinct products build on it, each with its own purpose.

**EMERGENCE: personal AI with deterministic memory.** EMERGENCE pairs a probabilistic reasoning model with OPAL's deterministic memory. The model generates answers; OPAL keeps the record and applies defined rules as time and accepted inputs change it. The design carries projects, relationships and corrections between sessions without a conventional RAG pipeline or a heavy vector database. Planned capability packages combine training examples, domain knowledge, tools and tests. A controlled learning process evaluates changes before accepting them. The research plan includes a C++ transformer runtime and new ways for model weights, the learned parameters, to interact with persistent memory. OPAL supplies the foundation. EMERGENCE develops the personal AI that the user owns and controls.

**KINDRED: models, tools and project memory.** KINDRED connects language models to project memory through its MCP server. Its application routes requests to the selected model and handles tool calls. Its public tools find project information, remember and correct accepted facts, and keep handoffs and whiteboard tasks. The assistant connects the selected model to memory and web tools. Projects retain their working record as users return to the task. Its Finder layer brings user-approved project files into that record with stable identities, change detection and traceable source excerpts. Section 7 reports the focused MCP checks and the distinct Finder evidence.

**PYAMV: the engine in Python.** PYAMV gives developers Python access to OPAL Atomic Memory Vertices, the native memory foundation. Developers can connect applications, call engine operations and run repeatable experiments beyond KINDRED's interface. Section 6 reports what the installed package did in our tests.

PYAMV also carries this work into robotics research. Our companion paper, *Temporal Observation Processing for Robotics*, reports preservation and new-process recovery of all 2,105 serialized observation records, including 301 records that retained explicit missing-reference values [12]. It explores noisy and incomplete measurements in separate synthetic experiments. Those preservation checks used JSON application content through a direct-build SDK module; they did not test the later installed wheel or native CBOR observation ingestion. The demonstrated capability is observation continuity across process restarts. Planning and physical control belong to the robotics application.

KINDRED and PYAMV support practical work and experiments. Their interfaces let people exercise the engine directly. EMERGENCE is the broader personal-AI research direction. OPAL, KINDRED and PYAMV supply the working engine and interfaces described here. EMERGENCE defines a separate personal-AI program.

## Availability and Free Local Experimentation

We plan to release KINDRED/MCP and PYAMV under free licenses for engineers and home builders to use and experiment with locally. We do not plan to charge for those releases. The goal is hands-on access: run the SDK workloads, inspect behavior, and build applications on the memory foundation. We built, installed and tested the macOS arm64 PYAMV package. We will provide the public download and precise license terms with the release. EMERGENCE remains a separate development and research program.

## Toward a Persistent Personal AI

The proposed EMERGENCE Personal AI (PAI) architecture gives this separation a concrete purpose [7]. A user-owned system could retain projects, accepted decisions, corrections, relationships and unfinished work across sessions. Returning to a project would begin from that record, with the assistant preparing the relevant context for the next useful action.

The engine tests establish inexpensive local updates, matching saved files and continuation after recovery. The application plan builds on that work: select useful information, let users inspect memory and check new information before accepting it. Those steps connect reliable storage to an assistant people can use.

The planned Context Compiler selects project information, decisions, sources and unfinished work for the current task. It keeps source history and distinguishes accepted information from temporary or disputed material. A temporary workspace holds the evidence while the task runs. A validation step checks what the system should keep. The proposed OPAL Finder lets users inspect and correct the stored record.

A compact context packet can reduce how much stored history the application sends to a model. The application must measure selection cost and task quality together: a small packet only helps if it contains what the task needs. The model can request additional evidence. A transformer still uses its active context and key-value (KV) cache, which holds intermediate attention data for generation. This design changes what information the application sends into that window.

Reinforcement, relevance, decay and relationships create a further opportunity: earlier experience could improve later retrieval and action. Memory records, approved training examples and model weights have separate roles in the PAI plan. Application tests must establish whether these mechanisms improve useful behavior over time.

### From Personal Systems to Server-Hosted Agents

The same separation applies to personal AI and server-hosted agents. Both need to keep records between model calls and supply useful information when work resumes.

In a data-center application, a continuity service could retain agent and project state between model requests, then prepare relevant context when reasoning resumes. The roles stay the same at a different scale. A service must also separate customers' data, order updates across machines, remain available and recover from failures. OPAL supplies local memory operations and recovery; the hosted application owns those service-level responsibilities.

### Local and Disconnected Work

Robotics, UAVs, distributed sensing and disconnected local AI also need continuity when connections offer little bandwidth. Local state operations can continue without a cloud connection. Reasoning during disconnection requires an available local model or waits for connectivity. The original iPhone simulator integration demonstrates an application connection; the current engine studies establish specific state-management rules.

Across these directions, OPAL ONE's flexibility comes from separating durable state from reasoning. Each application chooses how to select context, call models and act on results. That separation lets the paper's central idea extend beyond any one assistant while keeping the measured engine boundary clear.

## 11. What People Can Do With It

OPAL and its interfaces are built. The engine performs local transitions within a five-million-node collection and passes its state, replay and conflict checks. PYAMV delivers those operations through an installable Python package. KINDRED puts persistent memory into project work. These are the engineering results reported in this paper.

Our planned free licensed releases put those capabilities in the hands of engineers and home builders. People can build local tools, persistent assistants and applications that carry work across interruptions. EMERGENCE takes the architecture toward personal AI under the user's control. That separate program builds on OPAL; it does not define unfinished work in this engine project.

Three application directions show how people can build on what we have delivered.

## Make Continuity Visible

An assistant can keep accepted decisions, relationships and unfinished steps, then recover that record when its process restarts. A public demonstration can make that continuity visible: stop the application, restart it and show the work it retains. The engine supplies persistence and recovery. The application controls external actions and their retry rules.

## Compare the Operating Advantage

Developers can compare two continuity strategies in the same application, using the same inputs, reasoning model and hardware. Useful outcomes, model calls, context preparation, CPU work, energy and recovery show the consequences at application level. Direct temperature and power measurements address the thermal question separately from the operation timings in this paper.

## Put Earlier Experience to Work

Applications can use reinforcement, relevance, decay and relationships to change what they retrieve as a project develops. A useful comparison tracks recall, corrections and completed work across repeated interactions. This evaluates the application benefit of evolving memory while keeping the engine's demonstrated state-management results distinct.

These are uses and extensions of the working engine. The present results establish its measured local operations, preservation and recovery, together with tested developer and project interfaces. Application comparisons examine the additional benefits people achieve with those capabilities.

Let models interpret, plan and write. Let OPAL carry the record between those sessions. We built it so people can build with it.

# 12. Scope of the Evidence

The evidence supports the named operations on the recorded implementation and hardware. The original 1M-25M observations, damaging-input findings, CPU report and density estimates remain historical material; the original controlled comparison covers 10k-1M, the 23 October follow-up extends that fixture to 5M, and the separately preserved older scale series covers 1M/5M. Those workloads are not interchangeable.

**CPU and temperature.** “Cold CPU” refers to the reported cool-running sustained operation, not a cold-cache benchmark. The new studies collected no temperature or power series. CPU utilization alone does not measure either. The operating observation motivated the ICE-AGE name; the new experiments measure work and correctness.

**Timing and access.** The main tests fix record shape, policy and context, prepare the indexes, and start with an empty traversal cache. Hash access carries expected or amortized bounds rather than a strict worst-case guarantee for every call. First calls, container growth, global maintenance, policy history, contention and paging add costs. The timer resolution and 15 subsequent receipt samples per process do not characterize rare slow calls in production. The 5M extension uses the same executables in a later shared-machine session with swap already in use, so host conditions and scale both affect interpretation.

**Reliability and state.** State-preservation, replay, concurrency and recovery checks support the rules they test. They do not estimate a field failure rate or guarantee unlimited operation. They also do not prove convergence for arbitrary input. The production update uses two coordinates; the full projected BLGC state differs. The companion manuscript explains that mathematical distinction without changing the engine's behavior.

The historical disruption tests report graph structure and replay outcomes under their stated workloads. Matching replay hashes establish preservation of the covered fields; they do not measure retrieval quality or prove that cleanup restores every useful relationship.

**Capacity and applications.** The reported work covers OPAL and the KINDRED/MCP and PYAMV interfaces described here. The complete PAI Context Compiler, personal memory application and distributed hosted service belong to separate application programs. The billion-node figures project file-size density; they are not deployment measurements. This study measures neither recovery across machines nor the effects of lower-precision arithmetic, retrieval quality or energy savings. Our evidence packages retain source snapshots, executable identities, samples and exact file comparisons. The paper reports their results while keeping private implementation details in those records.

These boundaries define the evidence, not a list of unfinished engine work: the scale runs passed their correctness checks, and separate version comparisons preserved exact saved state and continuation after recovery.

# 13. Conclusion

OPAL keeps the record while reasoning pauses. It retains identities and relationships, applies accepted updates, and recovers saved state so work can continue. We built the engine, put it behind practical interfaces and tested what it does. BLGC supplies the mathematics. The engine, wrappers and experiments supply the engineering evidence.

The current engine records a resident-update P50 of 0.083 microseconds at 1M, 5M and 1M in one sequential session. All three processes pass their correctness checks. The earlier five-run 5M series records subsequent receipt medians of 4.708-5.166 microseconds on its recorded version. The experiments establish their respective operation costs; separate tests expose storage-growth and deliberate-miss costs.

Reliability supports that result through its own checks. Independent processes recover saved state, recognize replayed identities and continue. Rejection, conflict, concurrent-identity and allocation-failure tests pass. Canonical checkpoint and continuation comparisons establish preservation across the versions documented in Appendix B.

KINDRED puts that foundation to work in projects. Its MCP interface carries facts, corrections, handoffs and tasks, while Finder brings user-approved files into the record with traceable sources. All 22 unique focused MCP checks pass, including project access, retry handling, correction history and whiteboard recovery. The application gives continuity a practical job.

PYAMV puts the same foundation in developers' hands. Its installed Python package passes 17 SDK contract checks and all four workload checks in each of five repetitions. Developers can retrieve records, add states and relationships, follow connected paths and recall identities through the public interface. The package turns engine capabilities into operations people can build with.

We plan to release KINDRED/MCP and PYAMV free for local experimentation by engineers and home builders. We built these interfaces to use the engine, test it and open the way for others to build on it.

The historical record adds sustained cool CPU operation near 17.2% utilization, successful replay under noisy input and gradual degradation under paging. The controlled tests show the work each operation performs and the state it preserves. The paper reports these operating observations separately from its controlled latency and correctness measurements.

What this means is straightforward: an application can keep its accepted record while the reasoning model is idle. That record has identities, relationships and defined update rules; the application can find, correct, save and recover it. KINDRED and PYAMV turn those capabilities into interfaces for model-assisted work and developer experimentation.

What comes next is wider use. We plan to release those interfaces free for engineers and home builders, then build on the same foundation in EMERGENCE: probabilistic reasoning with deterministic memory, controlled learning and capability packages under the user's authority. Hosted agents offer another direction for the architecture. Each application builds on the engine and interfaces demonstrated here.

That separation creates the Compute ICE-AGE opportunity. OPAL keeps the accepted record. The model reasons over it when the task calls for reasoning. We built the engine. KINDRED uses it. PYAMV opens it to developers.

The system did not eliminate computation. It changed where computation occurs.

# Author’s Note

As I near the end of these revisions, I want to return to the purpose that started this work. I developed this memory engine for edge devices, with a simple conviction: people should have the power to decide where their memory lives, who receives their data, and what systems may act on it. I wanted to move that authority closer to the person.

Through many builds, the architecture revealed more than I initially set out to find. I observed cool-running operation, deterministic state preservation, and local traversal behavior that remained stable as the graph grew within the tested workloads. Those findings widened the architectural ambition - from memory on a personal device to continuity services in a data center - while preserving the original purpose.

**Scale creates possibilities. Ownership gives them direction.**

Today, the possibility that most captures my imagination is a decentralized personal AI: a system that carries our projects, decisions, and experiences forward under our control. Its reasoning may pause. Its continuity can endure. Its memory should remain ours to inspect, correct, protect, and share.

That possibility will form the foundation of my next paper. This work establishes the ground beneath it.

Thank you for taking the time to read my work.

*R. Jay Martin*

# References and evidence

[1] Martin, R. Jay II. (2026). *Bounded Local Generator Classes for Deterministic State Evolution*. arXiv:2602.11476v2, revised 7 May 2026. https://arxiv.org/abs/2602.11476v2. Public mathematical foundation; the entry below identifies the later author-held revision.

[2] Martin, R. Jay. (2026). *Bounded Local Generator Classes for Deterministic State Evolution*. Revised author-held manuscript, 22 October 2026, 42 pages. Banach-space formulation, production correspondence, reliability and controlled scaling evidence. We distinguish this revision from the version currently on arXiv.

[3] *BLGC v3 OPAL Benchmark Evidence* and preserved six-run 1M/5M scale records, October 2026. Metric definitions, source/runtime provenance, indexed-access and sampled-memory observations. Author-held evidence; consolidated in [2].

[4] *BLGC production generator study*, 22 October 2026. Nine initial fresh-process scale runs; frozen production source, raw samples, validation, independent-process recovery and private correspondence analysis. Evidence identity: blgc-generator-20261022.

[5] *Matched receipt-index repair study*, 22 October 2026. Baseline/repaired frozen sources, 18 matched scale processes, canonical differential files, correctness logs, nine forced-growth processes and raw samples. Evidence identity: blgc-receipt-index-20261022. This paper reports aggregate results; implementation details remain in private records.

[6] *Complete-query concurrency evidence*, October 2026. Four fresh runs of a 513-node, 512-candidate fixture; 32,000 queries across idle and concurrent-maintenance cases. Author-held records summarized in [2].

[7] *Emergence + OPAL: Personal AI Architecture and Build Plan*. Author-held architecture and build-plan memo, reviewed 22 October 2026. Source for the proposed Context Compiler, task workspace, inspectable memory and controlled learning architecture; not an application benchmark.

[8] *Compute ICE-AGE 5M controlled extension*, 23 October 2026. Ten sequential fresh processes, five baseline and five repaired; identical frozen binaries and fixture to [5], alternating version order. Raw samples, per-process correctness reports, binary hashes, execution order and host-condition note. Evidence identity: tcia-5m-20261023.

[9] *5M timing-boundary and index-work audit*, 23 October 2026. Six attribution processes across 1M/5M using frozen production libraries: natural pre-repair, separately prepared pre-repair, and natural repaired paths. Index-state/capacity observations, raw timing rows, independent expected-state checks and matching canonical receipt outcomes. Evidence identity: tcia-5m-audit-20261023. Diagnostic timings are separate from [8].

[10] *Installed-wheel PYAMV SDK validation*, 26 October 2026. Current-source Release wheel, clean Python 3.14.6 macOS arm64 installation, 17 SDK contract checks, five repetitions of four local Benchpress-compatible workloads, source and binary identities, raw reports and rate reconciliation. Evidence identity: tcia-sdk-20261026. Author-run SDK evidence; not upstream DCPerf or independent third-party certification.

[11] *KINDRED focused MCP repair and Finder evidence audit*, supplied 26 October 2026. Xcode RunSomeTests: 12 representative checks and ten additional checks passed; two repaired checks also passed independently, for 24 executions covering 22 unique tests. Xcode BuildProject completed with zero errors. Finder inventory: 40 tests, not run; historical focused passes: 16, 20, 22 and 25. Author-supplied audit report; source identities and repair scope appear in Appendix C. No full 455-test pass or independent certification claim.

[12] *Temporal Observation Processing for Robotics: PYAMV integration and early smoothing experiments*. Author-held working whitepaper, 18 October 2026, Sections 18–19. Direct-build SDK serialized-observation preservation, isolation, publication retry and new-process recovery; separate synthetic observation experiments. Native CBOR observation admission and physical-robot validation remain outside this evidence.

[13] *Current-engine resident-operation validation*, 26 October 2026. Frozen binary, same-session 1M-5M-1M sequence, raw samples, correctness and independent recovery checks, canonical checkpoint comparisons, and deliberate collision workload. Evidence identity: resident-hot-index-20261026. One pre-existing aggregate contract check fails on both compared versions; the report retains it separately from the passing focused checks.

# Appendix A. Preserved historical figures and metric context

This appendix retains the original figures and their embedded labels to preserve the historical argument and visual record. The figures depict the earlier account; the new controlled data appear in the main results tables.

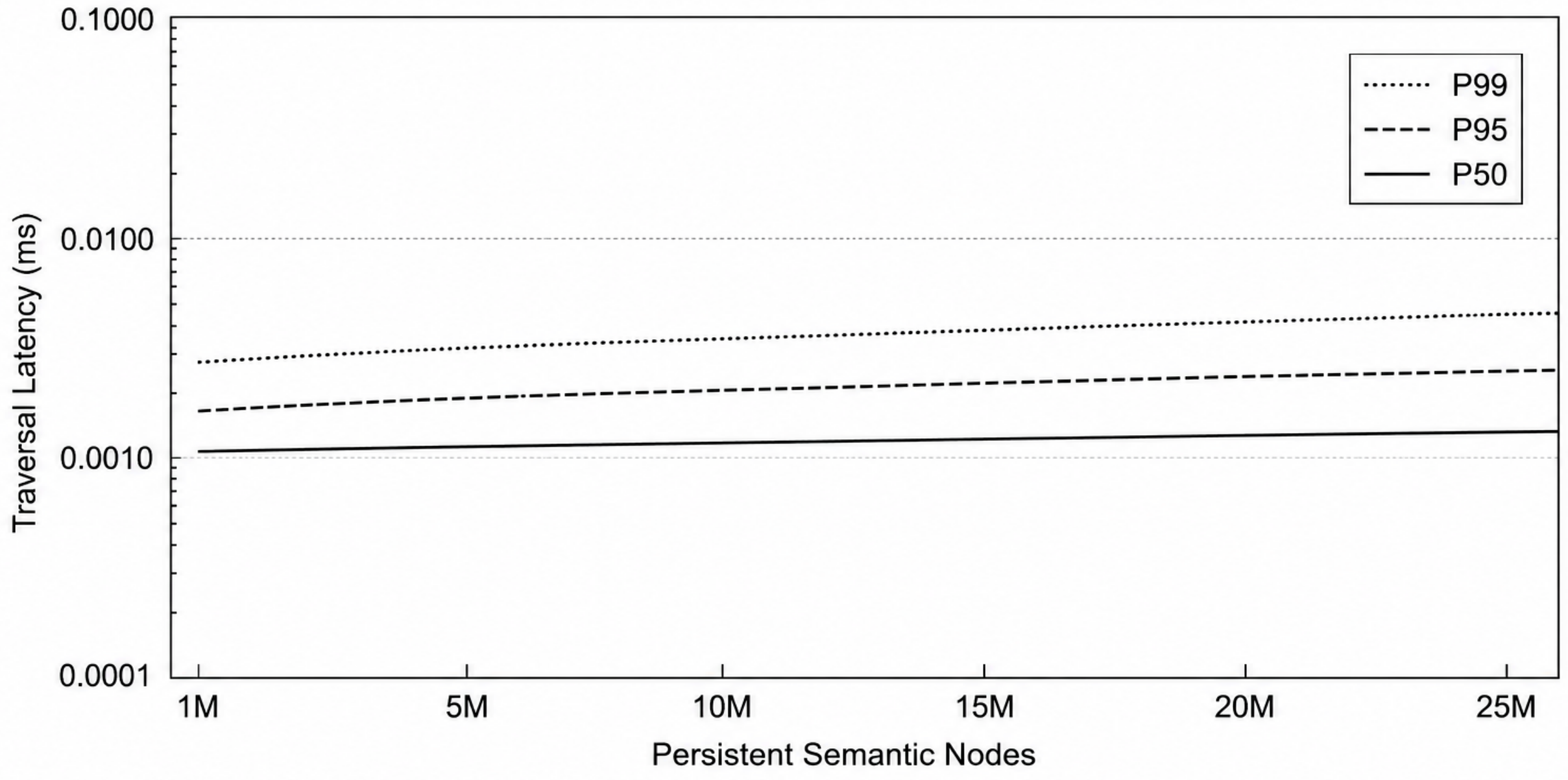

**Figure 1 — Traversal Stability Across Persistent Semantic Scaling Regimes**

*Traversal latency remained locality-constrained across scaling regimes ranging from 1 million to 25 million persistent semantic nodes. P50, P95, and P99 latency intervals widened gradually under increasing graph cardinality and paging pressure, but catastrophic traversal divergence was not observed. Operational work remained strongly coupled to bounded local traversal behavior rather than total persistent graph mass.*

**Figure 1. Original traversal-stability figure.** The figure depicts the earlier 1M-25M traversal account. The new study reports resident updates, passive snapshots and receipt transactions with explicit timing definitions; it does not independently validate every plotted historical point.

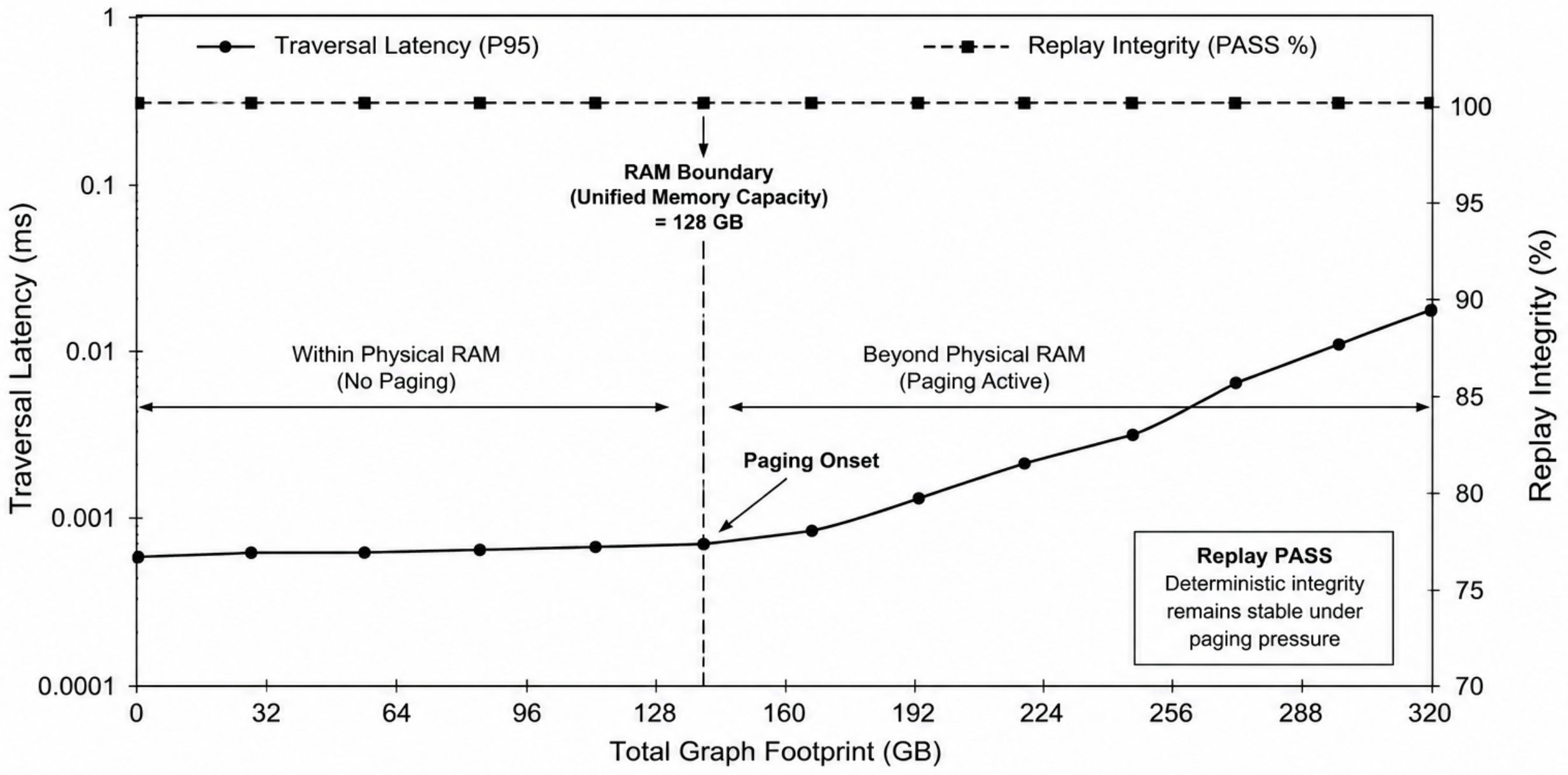

**Figure 2 — Memory Pressure and Paging Behavior Under Unified Memory Oversubscription**

*As total graph footprint exceeded available physical unified memory, active paging pressure increased traversal latency and locality fragmentation. However, degradation remained incremental rather than catastrophic. Deterministic replay integrity and bounded traversal continuity survived sustained paging conditions despite fragmented topology and stochastic perturbation workloads.*

**Figure 2. Original paging-pressure figure.** The embedded diagram labels a 128 GB RAM boundary and graph footprints up to 320 GB. Those labels do not describe the 16 GB M2 Pro environment used for the new experiments. We keep the historical illustration and do not claim it measures the paging curve for that machine.

## Metric reconciliation

The supplied abstract reported a traversal mean near 0.32 ms, while its body reported traversal P50 near 0.0014 ms. The supplied text does not provide a common raw-sample definition connecting those values, so this revision preserves them as historical labels rather than combining them with new timing tables. A separately preserved 5M indexed-access report labels 1.40751 μs as P95 across per-ID means [3]; that statistic measures indexed access, not a traversal median.

The present paper uses operation-specific definitions throughout: per-ID means for the older indexed-access series; individual-call quantiles for the controlled resident and receipt series; and separately forced capacity events. “Stable across scale” refers to the reported statistic and workload, not identical latency for every operation or the absence of paging/growth costs.

# Appendix B. Historical baseline, implementation equivalence and timing audit

The main text reports repaired-engine operation. This appendix preserves the pre-repair comparison and the evidence explaining the change. Baseline costs describe the former implementation, not the ongoing ordinary receipt cost of the repaired engine.

## Matched study design

The matched comparison built original and repaired source snapshots in native arm64 Release with Apple clang 21.0.0, C++17 and -O3 on macOS 26.6.2. Only the receipt-index maintenance differed. Eighteen fresh processes cover three repetitions at each of 10,000, 100,000 and 1,000,000 nodes for each version. The study alternates version order and counterbalances scale order. One caller performs the operations; OPAL_THREAD_COUNT is one. No profiler was active.

On 23 October, a separate extension repeats the same fixture at 5,000,000 nodes in ten sequential fresh processes: five baseline and five repaired. Version order alternates by repetition. Executable hashes match the frozen binaries used on 22 October; the extension changes node count and adds repetitions without changing engine code or correctness checks. Each version adds 102,400 passive snapshots, 102,400 resident updates and 80 receipt events. The host already reported 3,311.75 MiB of swap in use before execution. This is a shared-machine follow-up, not a new counterbalanced rerun of all scales [8].

## Historical Receipt Transactions

**Table B1.** Ranges of per-process P50, in microseconds: three repetitions per version at 10k-1M on 22 October. Receipt values are per-run P50s over calls 2-16; the text below reports the single first receipt observation from each run separately. Both receipt columns time the complete in-memory public event call. The pre-repair column is historical comparison data: it includes repeated full-index rebuilds that the repaired path removes. Neither column times checkpoint publication to disk.

| Data nodes | Pre-repair receipt P50, μs | Repaired receipt P50, μs |
|---|---|---|
| 10,000 | 422.167-454.708 | 5.166-6.459 |
| 100,000 | 8743.042-9897.792 | 4.875-5.459 |
| 1,000,000 | 177699.667-182297.375 | 4.458-4.708 |

At one million nodes, subsequent receipt-event P50 fell from 177.700-182.297 milliseconds to 4.458-4.708 microseconds. The original path invalidated the full ID index after each receipt append;

the repaired path maintains the new entry incrementally under the same lock. Existing record positions remain valid after append, even when vector storage moves. This removes repeated full-index reconstruction from the tested event path.

The first repaired receipt remained separately observable: 17.417-18.584 microseconds at 10k, 15.542-21.334 at 100k, and 19.333-23.333 at 1M. The 15 subsequent calls per run provide a bounded diagnostic, not a production tail estimate. The very short resident timings require the timer-resolution qualification: the initial nine-run study recorded empty timer-pair P50 of zero and P95 of 42 ns; a one-nanosecond median difference is not a meaningful speed change [4].

**At 5M, the earlier receipt series recorded subsequent receipt medians of 4.708-5.166 microseconds [8].** Current-engine resident-update results appear in Table 1.

Each run supplied one first receipt observation. At 5M, those observations ranged from 28.500 to 651.792 microseconds. The experiment did not isolate its cause, and the clean-index starting condition rules out attributing it to the documented post-append index rebuild. The runs took place in a later shared-machine session with preexisting swap use; the experiment did not isolate paging, scheduling, allocation or other first-use effects. We assign no cause or cardinality-scaling conclusion to these observations.

One first-call observation per process cannot define a latency distribution; the repeated-call summaries in Table B1 use calls 2-16 and do not estimate production tails.

Peak process RSS ranges from 3.541-4.604 GB for baseline and 4.377-4.638 GB for repaired runs (decimal GB, macOS high-water accounting). Existing system swap use does not establish how much paging occurred inside any timed call. The extension reconciles every reported timing statistic against raw samples and retains setup, index-preparation and timer diagnostics [8].

All ten 5M processes pass the original reference-state, record/topology, chronology, generation, passive-read, receipt-count, replay and conflict checks. The extension does not rerun the separate allocation-fault, canonical checkpoint comparison or forced-growth fixtures; their existing results retain their original scope.

## Audit of the 5M Timing Boundary

A separate attribution diagnostic tests the source-level explanation [9]. At 1M and 5M, it compares three cases: ordinary pre-repair calls; pre-repair calls after a separately timed public snapshot; and repaired calls. The snapshot rebuilds a dirty index before the next receipt enters its timer. The diagnostic observes index state and container capacity without changing engine code or canonical state rules.

At 5M, the ordinary pre-repair receipt median is 1.157 seconds. Moving the rebuild into the preceding snapshot moves the large cost with it: snapshot median 1.169 seconds, receipt median 25.458 microseconds. The repaired diagnostic median is 7.667 microseconds with a clean index before and after every event. No timed event grows the vector capacity or hash bucket count. All three paths produce identical checked numerical state and canonical receipt outcomes; replay, conflict, chronology and generation checks pass. These diagnostic timings explain the mechanism and do not replace the uninstrumented results in Table B1.

The timed call includes building the payload, acquiring the lock, finding and validating the target, calculating the update, appending the receipt, maintaining the index and making the change visible. Setup, initial index construction, warmup and post-run verification sit outside that boundary. Resident-update measurements exercise 1,024 prepared targets, not a scan or traversal of all five million records. Source inspection finds no retained-record scan in the resident path: it locks, resolves the ID through the derived hash index, accesses the resident record, evaluates the fixed numerical update and publishes generation state.

The audit confirms the repair's mechanism: the old path revisits retained records to reconstruct an index; the repaired path indexes the appended receipt. It also corrects a historical fixture label: the JSON field receipt_dirty_index means calls 2-16 in both versions. Only the pre-repair calls start with a dirty index. The paper reports those samples as subsequent receipt calls.

## State preservation through the change

The repair changes only how the engine maintains its derived index. Numerical evaluation, request admission, canonical receipt payloads, chronology, usage, publication and durable encoding keep their existing behavior. If the engine cannot allocate an index entry after appending a receipt, the helper marks the cache dirty and allows the existing canonical commit to finish. The next lookup rebuilds the index through the existing fallback.

# Appendix C. KINDRED audit scope and repair record

The audit identified two failures and corrected both. Universal Transport already rejected disabled or mismatched Project Access without backend execution, but the HTTP adapter mapped its denial payload to HTTP 200. A guard at the tools/call boundary now requires resolver approval, matching project binding, enabled Project Access and a valid binding revision; failure returns an empty HTTP 403. The hardening fixture used an obsolete client identity. The repair substituted the recognized Codex identity and strengthened empty-body assertions without changing the production allowlist.

The audit changed OpalMCPServer.swift and KINDREDTests.swift. It reports no OPAL C++ changes, persistence-law changes, authorization-policy changes or public-tool-schema changes. The six-tool registry check passed. Each check ran individually and sequentially through Xcode RunSomeTests, followed by BuildProject. Individual summaries, console logs and result bundles remain in that audit session's RunSomeTests directory. Source identities remain in the supporting audit record.

The audit also records unchanged shared-contract, transport-runtime and native transport identities. SourceEditor diagnostics timed out on the two large Swift files; targeted test builds and the final project build succeeded. An older MCP V1 document still says five tools; executable contract 1.1 and its tests establish the current six-tool interface. These focused results do not substitute for the unrun full application or current Finder suites.

## Focused results and Finder evidence

KINDRED gives OPAL a working role in model-assisted projects. Its six public MCP tools are context, recall, remember, correct, handoff and whiteboard. Focused validation passed all 12 representative MCP checks and ten additional authorization and transport checks [11]. The audit executed 24 tests covering 22 unique checks because it ran the two repaired tests twice. The final project build completed with zero errors.

These checks covered tool discovery, credential rejection, project binding, durable idempotent writes, correction history, handoffs, whiteboard isolation and recovery, request cleanup and diagnostic redaction. Disabled or mismatched Project Access returned an empty HTTP 403 before backend execution. An authorized matching-project request returned HTTP 200 and invoked the dispatcher exactly once. Transport checks also verified invalid-origin rejection, the one-megabyte request limit and daemon cleanup. Together, these results show how the application controls access to persistent project work.

The current Finder plan lists 40 tests. Earlier focused suites passed at 16, 20, 22 and 25-test milestones as the feature grew. The audit inventoried the current suite but did not run it or the full 455-test application plan. We therefore report current focused MCP results and historical Finder results separately. Finder supports identifier-text fallback and application-metadata relationships, rather than a complete parsed-symbol index or unrestricted native relationship-graph search. Appendix C records the repair and evidence identities.